\documentstyle[12pt,epsf,epsfig,wrapfig]{article}
\begin{document}
\def\b{\bar}
\def\d{\partial}
\def\D{\Delta}
\def\cD{{\cal D}}
\def\cK{{\cal K}}
\def\f{\varphi}
\def\g{\gamma}
\def\G{\Gamma}
\def\l{\lambda}
\def\L{\Lambda}
\def\M{{\Cal M}}
\def\m{\mu}
\def\n{\nu}
\def\p{\psi}
\def\q{\b q}
\def\r{\rho}
\def\t{\tau}
\def\x{\phi}
\def\X{\~\xi}
\def\~{\widetilde}
\def\h{\eta}
\def\bZ{\bar Z}
\def\cY{\bar Y}
\def\bY3{\bar Y_{,3}}
\def\Y3{Y_{,3}}
\def\z{\zeta}
\def\Z{{\b\zeta}}
\def\Y{{\bar Y}}
\def\cZ{{\bar Z}}
\def\`{\dot}
\def\be{\begin{equation}}
\def\ee{\end{equation}}
\def\bea{\begin{eqnarray}}
\def\eea{\end{eqnarray}}
\def\half{\frac{1}{2}}
\def\fn{\footnote}
\def\bh{black hole \ }
\def\cL{{\cal L}}
\def\cH{{\cal H}}
\def\cF{{\cal F}}
\def\cP{{\cal P}}
\def\cM{{\cal M}}
\def\olam{\stackrel{\circ}{\lambda}}
\def\oX{\stackrel{\circ}{X}}
\def\const{{\rm const.\ }}
\def\ik{ik}
\def\mn{{\mu\nu}}
\def\a{\alpha}
\begin{center}
{\bfseries Kerr Geometry as Space-Time Structure of the Dirac
Electron} \vskip 5mm A.Burinskii$^{1 \dag}$, \vskip 5mm {\small
(1) {\it NSI Russian Academy of Sciences}}
\end{center}
\vskip 4mm
\begin{abstract}
The combined Dirac-Kerr model of electron is suggested, in which
electron has extended space-time structure of Kerr geometry, and
the Dirac equation plays the role of a master equation controlling
polarization of the Kerr congruence. The source contains a
spinning disk bounded by a closed singular string of Compton size.
 It is conjectured that this Compton structure may be observed
for polarized electrons under a very soft coherent scattering.
\end{abstract}
\vskip 4mm {\bf Introduction.} The Kerr-Newman solution has
gyromagnetic ratio $g=2,$ as that of the Dirac electron (Carter
1968) and represents a classical model of extended electron in
general relativity (for references see [1,2,3,4]). There appears a
natural question, what is the relation between the Dirac equation
and the Kerr-Newman solution?
\par
The answer is related with the problem of coordinate description
of the Dirac electron which cannot be localized inside the Compton
region. Absence of the clear space-time description prevents the
consistent incorporation of gravity. Similar, in the
multi-particle QED theory, the ``dressed'' electron is smeared
over the Compton region. However, its coordinate description is
again very obscure and main results are obtained by calculations
in the momentum space. As a result, there appears some extreme
point of view that the subsequent relativistic theory has to
refuse from the wave function in coordinate representation at all
[5]. These facts hinder from natural incorporation of gravity. On
the other hand, ignorance of gravity in the Dirac theory and QED
is justified by  weeakness of the gravitational field of electron.
However, electron has the extremely large spin/mass ratio (about
$10^{44}$ in the units $\hbar=c=G=1 ,$) which shows that
gravitational effects have to be estimated on the base of the
Kerr-Newman solution. The extremely high spin leads to the very
strong polarization of space-time and to the corresponding very
strong deformation of electromagnetic (em-) field which has to be
aligned with the Kerr congruence. However, the {\it em-field of
electron cannot be considered as small}, and the resulting
influence turns out to be very essential! In particular, the
em-field turns out to be singular at the Kerr ring which is a
closed string of the Compton size $a=J/m.$ The space-time acquires
two folds with a branch line along the Kerr string and represents
a very non-trivial background for electromagnetic processes.
\par
In this work  we obtain an exact correspondence between the wave
function of the Dirac equation and the spinor (twistorial)
structure of the Kerr geometry. It allows us to assume that the
Kerr-Newman geometry reflects the specific space-time structure of
electron, and electron contains really the Kerr-Newman circular
string of Compton size. We suggest a combined Dirac-Kerr model of
an electron, in which electron acquires the coordinate description
of the Kerr geometry, while the Dirac equation plays the role of a
{\it master equation} which controls the position and dynamics of
the Kerr string and the related twistorial polarization of the
Kerr space-time. Dynamics of this combined Dirac-Kerr model in the
external electromagnetic fields turns out to be indistinguishable
from the behavior of the Dirac electron.
\par
{\bf Real structure of the Kerr geometry}
Angular momentum of electron $J=\hbar /2$  is extremely high with
respect to the mass, and the black hole {\it horizons disappear}
opening the naked Kerr singular ring which represents a closed string [3],
excitations of which generate spin and mass of the
extended particle-like object - ``microgeon'' [1]. Singular ring may be regularized
by Higgs field. If the Kerr string acquire tension $T,$  $m=E=Ta ,$ the Kerr relation $J=ma$ yields
 the Regge behavior $J=\frac 1T m^2.$
\par
{\it The Kerr principal null congruence} is a twisted family of
the lightlike rays -- twistors. Frame of the Kerr geometry is
formed by null vector field $k^\m (x)$ which is tangent to the
Kerr congruence. The Kerr-Schild form of metric is \be g^\mn
=\eta^\mn + 2H(x) k^\m k^n \label{KS}, \ee where $\eta^\mn$ is
metric of an auxiliary Minkowski space-time $M^4$ and $H$ is a
real function, $x^\m =(t,x,y,z).$

\begin{wrapfigure}[14]{R}{5cm}
\begin{center}
\mbox{\epsfig{figure=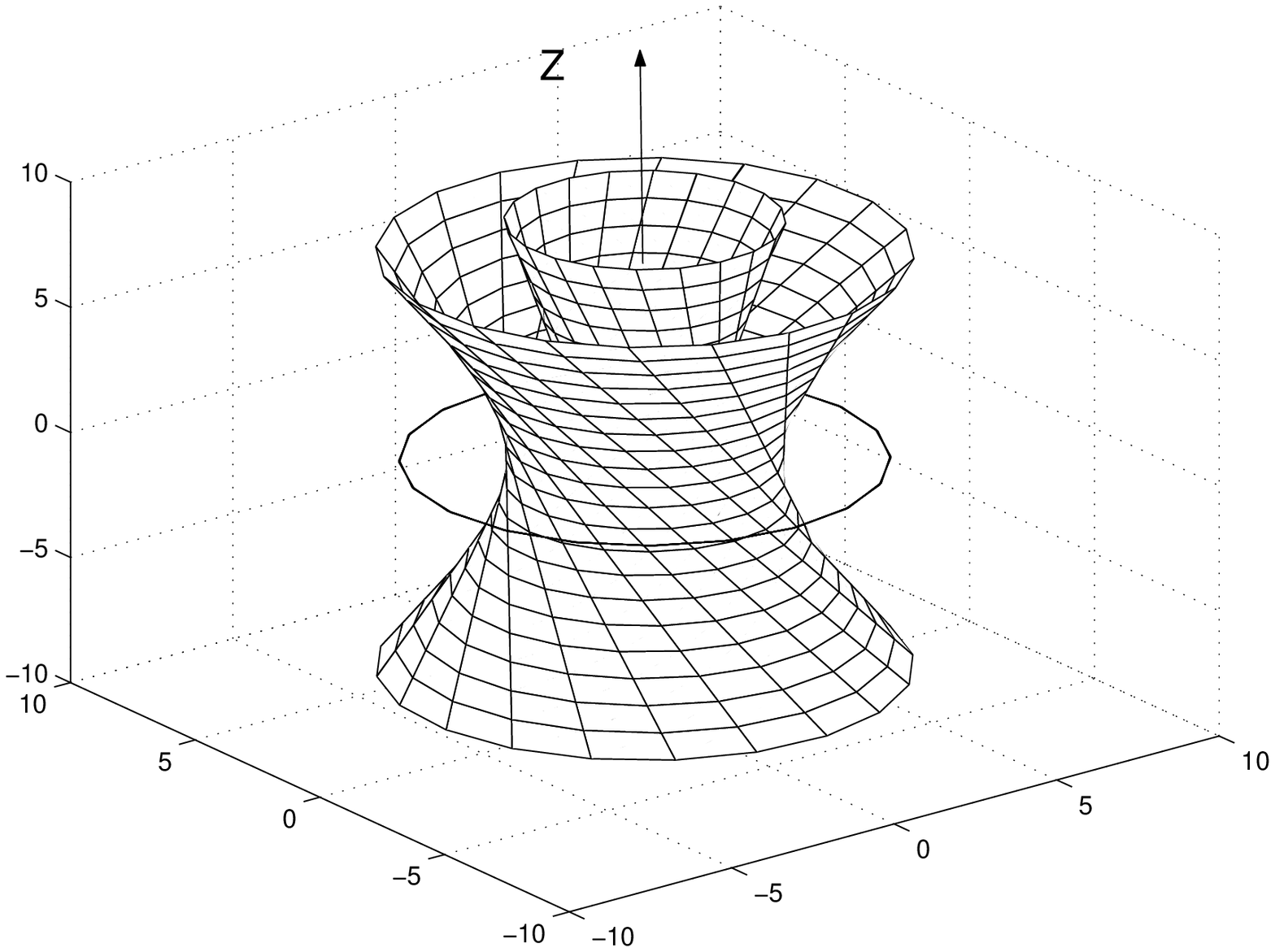,width=5cm,height=4cm}}
{\small{\bf Figure 1.} The Kerr singular ring and  congruence.}
\end{center}
\label{Burinskii_fig1}
\end{wrapfigure}

Vector potential of the Kerr-Newman solution is aligned with this
congruence \be A_\m = {\cal A} (x) k_\m \label{Aem},\ee and the
Kerr singular ring represents its caustic, see Fig.~1. {\bf The
Kerr theorem} determines the Kerr congruence via a holomorphic
surface in the projective twistor space which has coordinates
 \be Y,\quad \l_1 = \z - Y v, \quad \l_2 =u + Y \Z \ , \label{Tw}
\ee where $ 2^{1\over2}\z = x + i y ,\quad 2^{1\over2} \Z = x - i
y , \quad 2^{1\over2}u = z - t ,\quad 2^{1\over2}v = z + t $ are
the null Cartesian coordinates.
Such congruences lead to
solutions of the Einstein-Maxwell field equations with metric
(\ref{KS}) and em-field in the form (\ref{Aem}). Congruence of the
Kerr solution is built of the straight null generators, twistors,
which are (twisting) geodesic lines of photons.
Therefore, for any holomorphic function $F ,$ the solution $Y(x^\m)$
of the equation $F(Y,\l_1,\l_2)=0 $ determines congruence of null
lines by the 1-form \be e^3= du + \bar Y d \zeta + Y d \bar\zeta - Y
\bar Y dv \label{e3}  .\ee  The null vector field $k_\m dx^\m=
 P^{-1}e^3$ up to a normalizing factor $P .$
Coordinate $Y$ is a projective spinor $Y=\phi_2/\phi_1,$ and
in spinor form  $k_\m = \bar  \phi _{\dot\alpha} \bar\sigma
_\m^{\dot\alpha\alpha}  \phi _\alpha .$

{\bf Complex representation of  Kerr geometry.}
Complex source of Kerr geometry is obtained as a result of
complex shift of the `point-like' source of the Schwarzschild
solution written in the Kerr-Schild form.
Applying the complex shift $(x,y,z) \to (x,y,z+ia)$ to the
singular source $(x_0,y_0,z_0)=(0,0,0)$ of the Coulomb solution
$q/r$, Appel (in 1887 !) obtained the solution $ \phi(x,y,z)= \Re e
\ q/\tilde r, $
 where $\tilde r =\sqrt{x^2+y^2+(z-ia)^2}$ turns out to be
complex. On the real slice $(x,y,z)$, this solution acquires a
singular ring corresponding to $\tilde r=0.$ It has radius $a$ and
lies in the plane $z=0.$  The solution is conveniently described
in the oblate spheroidal coordinate system $r, \ \theta,$ where
$\tilde r =r+ia\cos\theta .$ The resulting real
space is twofold having positive sheet $r>0$,  and negative one $r<0$.

\begin{wrapfigure}[12]{R}{5cm}
\begin{center}
\mbox{\epsfig{figure=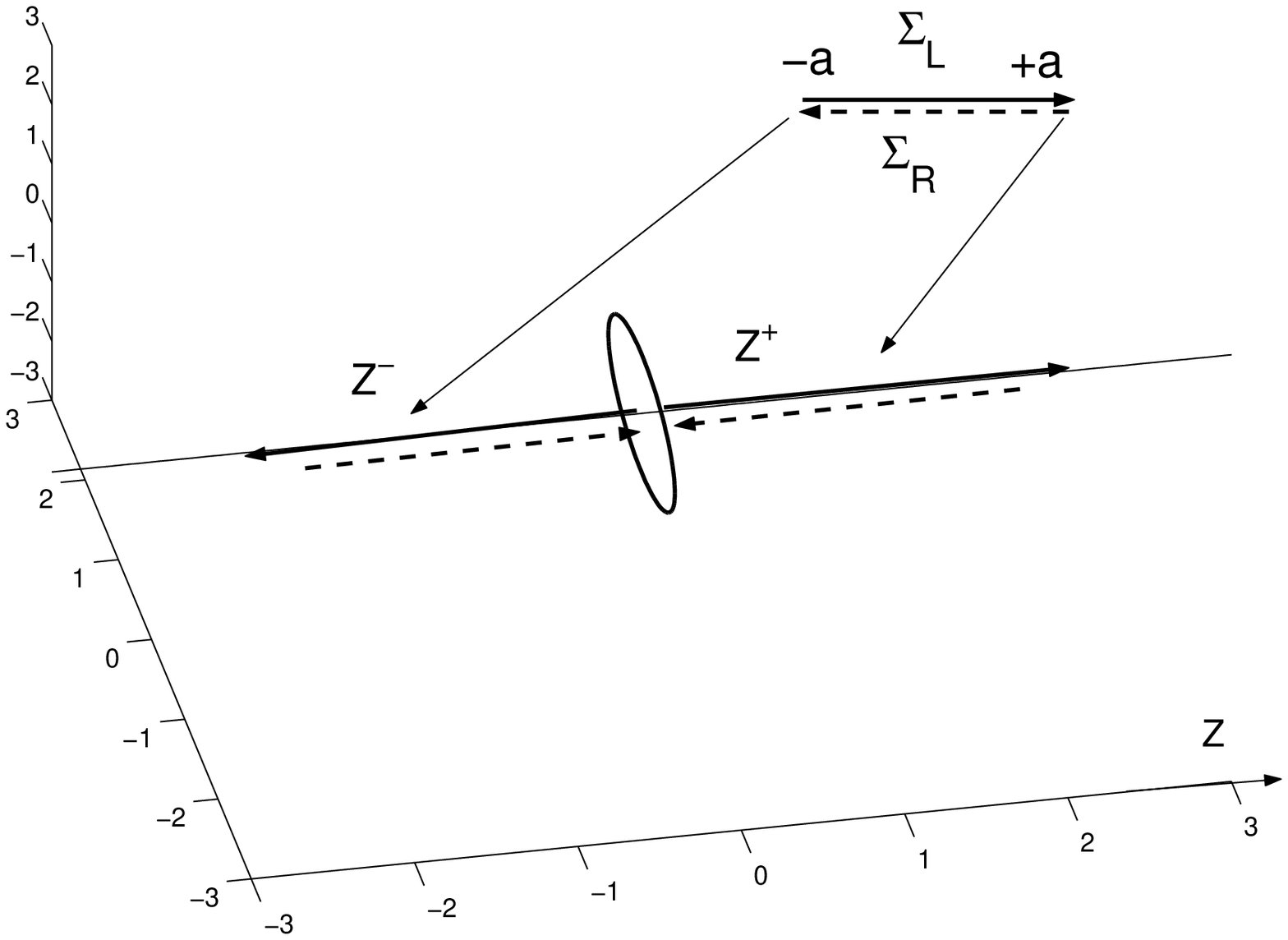,width=5cm,height=4cm}}
{\small{\bf Figure 2.} Singular ring and two singular half-strings.}
\end{center}
\label{Burinskii_fig2}
\end{wrapfigure}

The Appel potential corresponds  exactly to electromagnetic field
of the Kerr-Newman solution written in the Kerr-Schild form [1].
The vector of complex shift $\vec a$ shows angular momentum of the
Kerr solution $\vec J=m \vec a.$ Newman and Lind suggested a
description of the Kerr-Newman geometry  in the form of a
retarded-time construction, where its source is generated by a
complex point-like source, propagating along a {\it complex world
line} $X^\m(\t) \in CM^4$.
\par
In the rest frame of the Kerr particle, one can form two null
4-vectors $k_L=(1,0,0, 1)$ and $k_R=(1,0,0, -1),$
 and  represent the 3-vector of complex shift $i\vec a=i \Im m X^\m$ as the
  difference $i\vec a =\frac{ia}{2} \{ k_L -
k_R \}.$  The straight complex world line corresponding to a free
particle may be decomposed to the form $ X^\m(\t) = X^\m(0) + \t u^\m + \frac{ia}{2} \{ k_L - k_R \}
,$
where the time-like 4-vector of velocity $u^\m=(1,0,0,0)$
 can also be represented via vectors $k_L$ and $k_R$
 $ \ u^\m =\d _t \Re e X^\m(\t)=\frac 12 \{ k_L+k_R \}. $
 One can form two complex world lines related to the complex Kerr
source, \be X_{+}^\m(t+ia) = \Re e X^\m(\t) + ia k^\m_L , \quad
X_{-}^\m(t-ia) = \Re e X^\m(\t) - i a k^\m_R , \label{X+} \ee
which allows us to match the Kerr geometry to the solutions of the
Dirac equation.
{\bf Complex Kerr string.}
The complex world line $X^\m (\t)$ is parametrized by  complex
time  $\t=t+i\sigma$ and represents the world sheet of a very
specific string extended along imaginary time parameter
$\sigma \in[-a,a]$. The Kerr congruence, gravitational and em-
fields are obtained from this stringy source by a
retarded-time construction which is based on the complex null
cones, emanated from the worldsheet of this complex string
[3,4].  The complex retarded
time equation  $ \t = t - r  + ia \cos \theta  $
sets the relation $\sigma =a\cos \theta$ between the complex points
$X^\m (t,\sigma)$ and angular
directions $\theta$ of the real twistor lines. One sees that
this string is open with the end points
$\cos\theta= \pm 1$ which correspond to  $X_{\pm}^\m=X^\m(t\pm ia).$  By
analogue with the real strings, the end points may be attached
to quarks. The complex light cones adjoined to the end points have a {\it
real slice in the form of two especial twistors}  having
the discussed above null directions $k_L^\m$ and $k_R^\m $ which
determine momentum and spin-polarization of the Kerr solution.
These twistors form two half-strings of opposite
chirality, see Fig. 2.
\par
{\bf Chirons and excitations of the Kerr singular ring.}
The twistor coordinate $Y$ is also the projective angular
coordinate $ Y=e^{i\phi}\tan\theta $ covering the
celestial sphere, $Y\in CP^1 =S^2.$
 The exact Kerr-Schild solutions have em-field
 which is determined by arbitrary analytical function ${\cal A} (Y),$
in particular ${\cal A} =eY^{-n}.$ The simplest case $n=0$ gives
the Kerr-Newman solution. The case $n=1$ leads to an axial singular line along the
 positive semi-axis z. Due to factor $e^{i\phi},$ em-field of this
 solution has winding number n=1 around axial singularity. Since
 there is also pole at singular ring,  $\sim (r +ia \cos\theta)^{-1},$
 the em-field has also a winding of phase along the Kerr ring.
 Solution with $n=-1$ has opposite chirality and singular line along the
 negative semi-axis z. These elementary exact solutions (`chirons') have
 also the wave generalizations
 ${\cal A} =eY^{-n}e^{i\omega\t}$ acquiring the extra dependence from the
 complex retarded time $\t .$ The wave chirons are
 asymptotically exact in the low-frequency limit
and describe the waves propagating along the Kerr circular string
and induced waves along axial half-strings [3,4].  By  lorentz
boost the axial half-strings acquire modulation by de Broglie
periodicity [3,4].
\par
{\bf Dirac Equation in the Weyl Basis}
 In the Weyl basis Dirac spinor has the form
 $\Psi =
\left(\begin{array}{c}
 \phi _\alpha \\
\chi ^{\dot \alpha}
\end{array} \right),$
and the Dirac equation splits into \be\sigma ^\m _{\alpha \dot
\alpha} (i \d_\m  +e A_\m)
 \chi ^{\dot \alpha}=  m \phi _\alpha , \quad
 \bar\sigma ^{\m \dot\alpha \alpha} (i \d_\m  +e A_\m)
 \phi _{\alpha} =  m \chi ^{\dot \alpha}.\label{Wspl} \ee
The Dirac current $ J_\m = e (\bar \Psi \gamma _\m \Psi) = e
(\bar\chi  \sigma _\m  \chi + \bar\phi  \bar \sigma _\m  \phi ), $
can be represented as a sum of two lightlike components of
opposite chirality $ J^\m_{L} = e \bar\chi \sigma^\m \chi \ ,
\qquad J^\m_{R} = e \bar\phi \bar\sigma^\m \phi. $
The corresponding null vectors
\be k^\m_{L} = \bar\chi \sigma^\m \chi \ , \quad k^\m_{R} =
\bar\phi \bar\sigma^\m \phi , \label{kLR} \ee determine the
considered above directions of the lightlike half-strings. The
momentum of the Dirac electron is $p^\m = \frac m 2 (k^\m_{L} +
k^\m_{R}),$ and the vector of polarization of electron
 in the state with a definite projection of spin
on the axis of polarization  is $n^\m = \frac 12 (k^\m_{L} -
k^\m_{R}).$ In particular, in the rest frame and the axial
z-symmetry, we have $k_{L}=(1,\vec k_{L})=(1,0,0,1)$ and
$k_{R}=(1,\vec k_{R})=(1,0,0,-1),$ which gives
 $p^\m = m(1,0,0,0),$ and $n^\m =  (0,0,0,1),$
which corresponds to  transverse polarization of
electron, $\vec n \vec p =0 .$
 The Dirac wave wave function sets also
synchronization of the null tetrad in the surrounding space-time,
playing the role of an `order parameter'.
\par
{\bf Dirac Equation as a Master Equation Controlling
Twistorial Polarization.}
Em-field of the Kerr-Schild solutions $F_{\mn}$ is to be aligned
with the Kerr congruence, obeying the constraint $F_{\mn}k^\m =0.$
Therefore, twistorial structure of the Kerr-Schild solutions
determines strong polarization of the em field. In particular,
the elementary em-excitations on the Kerr background lead to  the
waves propagating along the Kerr circular string. Virtual photons are also
concentrated near this string, forming its excitation.
There is exact correspondence between two null
vectors (\ref{kLR}) obtained from the Dirac wave
function and similar vectors $k_L$ and $k_R$ related to the ends
of the complex Kerr string, Fig. 2. It allows us to
 unify the  Dirac and Kerr structures, considering
the Dirac equation as a master equation controlling twistorial polarization of the
Kerr space-time.
\par
{\bf Scattering.} Contradiction between the discussed Compton size
of electron and the results obtained for the deep inelastic
scattering is seeming and has simple explanation. Relativistic
boosts lead to asymmetry: $p_L << p_R$ or $p_L>>p_R$ which
determines the sign of helicity. As a result one of the axial
half-strings turns out to be strongly dominant.
 It allows one to use  perturbative twistor-string model [6,7]
which is based on  a reduced description in terms of the lightlike
momentum and helicity, and amplitude of scattering is determined
{\it only by  one} of the axial half-strings. We conjecture that
{\it the Compton size of the Dirac-Kerr electron may be observed
for polarized electrons under a very soft resonance scattering.}

{\bf References.}

[1] A.Ya. Burinskii, Sov.Phys. JETP, {\bf39} 193 (1974); Sov.Phys.
J. {\bf17} 1068 (1974).

[2] A.Ya. Burinskii, Phys. \ Lett. {\bf A 185} 441 (1994); gr-qc/9303003.

[3] A. Burinskii, Phys. \ Rev. D {\bf 70}, 086006 (2004), hep-th/0507109, 0710.4249 [hep-th].

[4] A. Burinskii, Grav.\&Cosmology, {\bf 10}, 50 (2004), hep-th/0403212.

[5] V.B. Berestetsky, E.M. Lifshitz, L.P. Pitaevsky, ``Quantum
Electrodynamics'', Oxford, UK:
Pergamon ( 1982).

[6] V.P. Nair, Phys. \ Lett. \ {\bf B 214} 215 (1988).

[7]  E. Witten, Comm. Math. Phys. {\bf 252}, 189 (2004), hep-th/0312171.
\end{document}